\documentclass{ws-ijmpe}

\begin{document}

\markboth{Su Houng Lee}{Chiral and $U_A(1)$ symmetry in correlation functions in medium}

\catchline{}{}{}{}{}

\title{CHIRAL  AND $U_A(1)$ SYMMETRY IN CORRELATION FUNCTIONS IN MEDIUM}

\author{\footnotesize Su Houng Lee}

\address{Department of Physics and Institute of Physcis \& Applied Physics, Yonsei University, 50 Yonsei-ro, Seodaemun-gu, Seoul 120-749, Korea\\
suhoung@yonsei.ac.kr}

\author{\footnotesize Sungtae Cho}

\address{Department of Physics and Institute of Physcis \& Applied Physics, Yonsei University, 50 Yonsei-ro, Seodaemun-gu, Seoul 120-749, Korea\\
sungtae.cho@yonsei.ac.kr}

\maketitle

\begin{history}
\received{(received date)}
\revised{(revised date)}
\end{history}

\begin{abstract}
In this review, we will discuss how the chiral symmetry and $U_A(1)$ breaking effects are reflected in the correlation functions. Using the Banks-Casher formula, one can identify the density of zero eigenvalues to be the common ingredient that governs the chiral symmetry breaking in correlation functions between currents composed of light quarks with or without a heavy quark.  Similarly the presence of the $U_A(1)$ breaking effect is determined through the contribution of the topologically non trivial configurations that depends on the number of flavors.  We also discuss how the symmetry breaking effects are reflected in the gluon correlation functions.  Finally, we review the Witten Veneziano formula for the $\eta'$ mass in medium.
\end{abstract}

\section{Introduction}

The breaking and its possible restoration of a symmetry at finite temperature and/or density is a fascinating subject as the possible  restoration in medium could be probed in relativistic heavy ion collision and/or in nuclear target experiments.
The restoration of chiral symmetry has been linked to the vector meson spectral density and has been the subject of great theoretical and experimental interest up to this day\cite{Hayano:2008vn,Leupold:2009kz,Rapp:1999ej}.  It has also been linked to quenching of the pion decay constant\cite{Jido:2008bk} and possible observation of the sigma meson in nuclear matter through the $\pi-\pi$ correlation\cite{Hatsuda:1985eb,Messchendorp:2002au}.
As for the $U_A(1)$ symmetry, its breaking is due to the anomaly, which induces an operator relation that remains broken above the QCD phase transition.  However, the physical effect, such as the large $\eta'$ mass, is intricately related to chiral symmetry breaking and the question of whether the mass will remain constant near the chiral symmetry restoration point is of particular interest as the partial quenching could be observed in nuclear target experiments\cite{Nagahiro:2008rj,Jido:2011pq,Nagahiro:2011fi,Nanova:2012vw}.
Moreover, the two pion Bose-Einstein correlation observed at RHIC seems to suggest the quenching of the $\eta'$ mass at high temperature\cite{Csorgo:2009pa,Vertesi:2009wf,Vertesi:2009wf-2}.

In this work, we will deliver a unifying picture on how the symmetries are reflected on the correlation functions.  In particular, we show through the Banks-Casher formula\cite{BC} that one can identify the density of zero eigenvalues to be the common factor that governs the chiral symmetry in all order parameters constructed from the  correlation functions\cite{Cohen:1996ng}.  Similarly the presence of the $U_A(1)$ breaking effect is determined through the contribution of the topologically non trivial configurations that depends on the number of flavors\cite{Lee:1996zy,Evans:1996wf}.  We also discuss how the symmetry breaking effects are reflected in the gluon correlation functions\cite{Kwon:2012vb}.  Furthermore, we show how these results can be used to generalize the Witten Veneziano formula to predict the properties of  $\eta'$ mass in medium\cite{Kwon:2012vb}.

\section{Chiral symmetry and zero eigenvalues}

\subsection{Banks-Casher formula}

The expectation value of an operator in QCD is defined as follows:
\begin{eqnarray}
\langle {\cal O}(x) \rangle & = & \frac{1}{Z} \int [d \mu] {\cal O}(x)
\end{eqnarray}
where the positive definite measure is defined as
\begin{eqnarray}
Z & = & \int [dA_\mu(x)] \exp [- \frac{1}{2g^2} \int d^4x F^2] {\rm det} [D \hspace*{-.2cm}\slash +m] \equiv   \int [d \mu] .  \label{measure0}
\end{eqnarray}
The notations and definition of the Euclidean path integral are introduced in the appendix.

It has been shown that from a Euclidean path integral point of view, the origin of chiral symmetry breaking is the presence of zero eigenvalues for the Dirac equation in the presence of the gauge fields\cite{BC}.

\begin{eqnarray}
\langle \bar{q} q(0) \rangle & = &
- \langle {\rm Tr} [ ( 0| \frac{1}{D \hspace*{-.2cm}\slash +m }|0) ] \rangle  = - \pi \langle {\rm Tr}[ J_{\lambda=0}(0,0)] \rangle  \nonumber \\
& = & - \pi  \langle  \int   \frac{ d^4x}{V} {\rm Tr}[ J_{\lambda=0}(x,x) ] \rangle  =-\pi \langle \rho(\lambda=0) \rangle,
\label{BC-0}
\end{eqnarray}
where we define a current density matrix of zero eigenvalues as follows:
\begin{eqnarray}
J_{\lambda=0}(x,y) & = & \sum_{\lambda=0}  \psi_{\lambda}(x) \psi^\dagger_{\lambda}(y).  \label{current-density-matrix}
\end{eqnarray}
Here,  $ i D \hspace*{-.2cm}\slash \psi_\lambda=\lambda \psi_\lambda$, and  the sum is over zero eigenvalues only.   $\rho(\lambda)$ is the density of eigenvalue $\lambda$.
Relation Eq.~(\ref{BC-0}) shows that the  essence of chiral symmetry breaking is the non-vanishing zero mode current density matrix $J_{\lambda=0}(x,y)$, which reduces to the scalar  density of zero eigenvalues $\rho(\lambda=0)$\cite{BC,Cohen:1996ng} in Eq.~(\ref{BC-0}).

\subsection{Correlation functions of light meson}

\subsubsection{Connected diagrams}

There are other operators and/or correlation functions that reflect the symmetry structure of the vaccuum/medium.  One can show that the presence of non-vanishing density of zero eigenvalues is the unifying ingredient that dictates the breaking of chiral symmetry in these operators.
To show this, let us  study the differences between meson two point functions of chiral partners.  The simplest example is the difference between scalar-pseudoscalar two point functions.
\begin{eqnarray}
\Delta_{S-P}^m (q) & = & \int d^4x e^{iqx} \langle \mathcal{T} \bigg( \bar{q} \tau^a q(x) \bar{q} \tau^a q(0)- \bar{q} \tau^a i\gamma^5 q(x)  \bar{q} \tau^a i\gamma^5 q(0) \bigg) \rangle  \nonumber \\
& = & \int d^4x  e^{iqx} \langle {\rm Tr}  [ \tau^a \tau^a \bigg( i \gamma^5 S(x,0) i \gamma^5 S(0,x)-S(x,0) S(0,x)   \bigg) ] \rangle  \nonumber \\
& = & \int d^4x e^{iqx} \langle {\rm Tr}  \bigg[- \tau^2 ( 0| \frac{1}{D \hspace*{-.2cm}\slash +m }|x)  \nonumber \\
&& \times \bigg(
( x| \frac{1}{D \hspace*{-.2cm}\slash +m }|0) -i \gamma^5 ( x| \frac{1}{D \hspace*{-.2cm}\slash +m }|0)  i \gamma^5 \bigg) \bigg] \rangle  \nonumber \\
 & \stackrel{m\rightarrow 0}{\longrightarrow} &
\int d^4x e^{iqx} \langle {\rm Tr}  \bigg[- \tau^2 ( 0| \frac{1}{D \hspace*{-.2cm}\slash +m }|x)\times J_{0}(x,0)  \bigg( 2 \pi  \bigg)
\bigg] \rangle \label{p-s-0}
\end{eqnarray}
Therefore, the difference vanishes when the current density matrix  of zero modes $J_0$  vanishes as in Eq.~(\ref{BC-0}). It should be noted that the argument of the wave function are identical in Eq.~(\ref{BC-0}), while it is not so in Eq.~(\ref{p-s-0}), the difference in $x$ will introduce a correlation length that is  not of interest at this point.

The difference between the vector and axial vector also has the common factor $(S(x,y)-i\gamma^5 S(x,y) i  \gamma^5)$  given in Eq.~(\ref{BC-identity4}), which is proportional to the density matrix of zero modes.  Eq.~(\ref{BC-identity4}) itself is gauge variant, but
becomes gauge invariant in the correlation function.
\begin{eqnarray}
\Delta_{V-A}^m (q) & = & \int d^4x e^{iqx} \langle \mathcal{T} \bigg( \bar{q} \tau^a \gamma_\mu q(x) \bar{q} \tau^a \gamma_\mu q(0)- \bar{q} \tau^a \gamma^5 \gamma_\mu  q(x)  \bar{q} \tau^a \gamma^5 \gamma_\mu q(0) \bigg) \rangle  \nonumber \\
& = & \int d^4x  e^{iqx} \langle {\rm Tr}  [ \tau^2 \bigg(  \gamma^5 \gamma_\mu S(x,0)  \gamma^5 \gamma_\mu S(0,x)-\gamma_\mu S(x,0) \gamma_\mu S(0,x)   \bigg) ] \rangle  \nonumber \\
 & \stackrel{m\rightarrow 0}{\longrightarrow} &
\int d^4x e^{iqx} \langle {\rm Tr}  \bigg[- \tau^2  \gamma_\mu ( 0| \frac{1}{D \hspace*{-.2cm}\slash +m }|x) \gamma_\mu  J_0(x,0)  \bigg( 2 \pi  \bigg)
\bigg] \rangle.
\label{v-a}
\end{eqnarray}
As with Eq.~(\ref{p-s-0}), Eq.~(\ref{v-a}) is non-vanishing when $J_0 \neq 0$ and so are all order parameters of chiral symmetry breaking.

\subsubsection{Disconnected diagrams}

If there is no flavor matrix in the scalar part of  Eq.~(\ref{p-s-0}), such as in the  SU(2) case, disconnected diagrams will contribute to the correlation function.
\begin{eqnarray}
\Delta_{S-P}^m (q) & = & \int d^4x e^{iqx} \langle \mathcal{T} \bigg( \bar{q}  q(x) \bar{q} q(0)- \bar{q} \tau^a i\gamma^5 q(x)  \bar{q} \tau^a i\gamma^5 q(0) \bigg) \rangle  \nonumber \\
& = & \int d^4x  e^{iqx} \langle {\rm Tr}  [ S(x,x) ] {\rm Tr}  [ S(0,0) ] \rangle + {\rm connected~terms}   \nonumber \\
& = & \int d^4x e^{iqx} \langle {\rm Tr}  \bigg[( x| \frac{1}{D \hspace*{-.2cm}\slash +m }|x) \bigg]  {\rm Tr}  \bigg[
( 0| \frac{1}{D \hspace*{-.2cm}\slash +m }|0)  \bigg] \rangle + \nonumber \\
 & \stackrel{m\rightarrow 0}{\longrightarrow} &
\int d^4x e^{iqx} \langle {\rm Tr}  \bigg[\pi J_0(x,x) \bigg]  {\rm Tr}
\bigg[\pi J_0(0,0) \bigg] + . \nonumber \\
\label{diss-cont}
\end{eqnarray}
Hence, the disconnected part also depends on $J_0$\cite{Cohen:1996ng}.

\subsection{Correlation functions of Baryons}

For the case of a nucleon, a typical interpolating field couples to both the positive and negative parity nucleon.  Several methods have been developed and attempted to isolate the parity eigenstates\cite{Jido:1996ia,Lee:1996au}.  The problem becomes simple, if one is interested in the differences in parity partners of the nucleons.  Here, one can look at few examples of differences between currents of chiral partners of the nucleon.  For example, an interpolating current for  $\Lambda$ could be $\eta_\Lambda = \epsilon^{abc} (u_a^T i \gamma^5 C d_b) h_c$; where $h$ is a heavy quark.  Then, we have
\begin{eqnarray}
\Delta_{\Lambda-\Lambda*}^m (q) & = & \int d^4x e^{iqx} \epsilon^{abc} \epsilon^{a'b'c'} \langle \mathcal{T} \bigg( (u^T_a i\gamma^5 Cd_b)h_c(x)  (\bar{u}_{a'} i\gamma^5 C^T\bar{d}_{b'}^T) \bar{h}_{c'}(0)
\nonumber \\
&& -  (u^T_a  Cd_c)h_c(x) (\bar{u}_{a'} C\bar{d}_{b'}^T) \bar{h}_{c'}(0)  \bigg) \rangle  \nonumber \\
& = & \int d^4x  e^{iqx}6  \langle {\rm Tr}  [S_d(x,0) C^T S_u^T(x,0)C -  S_d(x,0) i \gamma_5 C^T S_u^T(x,0) C i \gamma^5  ] S_h(x,0) \rangle  \nonumber \\
& = & \int d^4x e^{iqx} 6 \langle {\rm Tr}  \bigg[ ( x| \frac{1}{-D \hspace*{-.2cm}\slash +m }|0)  \bigg(
( x| \frac{1}{D \hspace*{-.2cm}\slash +m }|0) -i \gamma^5 ( x| \frac{1}{D \hspace*{-.2cm}\slash +m }|0)  i \gamma^5
  \bigg) \bigg] S_h(x,0) \rangle  \nonumber \\
 & \stackrel{m\rightarrow 0}{\longrightarrow} &
\int d^4x e^{iqx} 12 \pi  \langle {\rm Tr}  \bigg[ S^\dagger (0,x)  J_{0}(x,0)
\bigg] S_h(x,0) \rangle,
\label{l-ls}
\end{eqnarray}
where the chiral limit is taken with the light quark mass.
Again, one notes that the difference is proportional to $J_0$.

When one replaces the heavy quark $h$ to the $u$ quark in
the first two lines in Eq.~(\ref{l-ls}) as in the case of the proton correlation function, there is an additional contraction contributing to the correlation function.  This extra term is given as follows:
\begin{eqnarray}
\Delta_{p-p*}^m (q) & = &  \int d^4x  e^{iqx}6  \langle  S_u(x,0) \bigg(
( x| \frac{1}{-D \hspace*{-.2cm}\slash +m }|0) -i \gamma^5 ( x| \frac{1}{-D \hspace*{-.2cm}\slash +m }|0)  i \gamma^5
  \bigg) S_u(x,0)    \rangle  \nonumber \\
 & \stackrel{m\rightarrow 0}{\longrightarrow} &
\int d^4x e^{iqx} 12 \pi  \langle    S_u (x,0)  J_{0}(x,0) S_u(x,0) \rangle,
\label{l-ls-1}
\end{eqnarray}
which is again proportional to $J_0$.

One can choose another form for the current $\eta_\Lambda = \epsilon^{abc} (u_a^T  C h_b) i \gamma^5 d_c$.  Then, the difference between chiral partners takes the following form:
\begin{eqnarray}
\Delta_{\Lambda-\Lambda*}^m (q) & = & \int d^4x e^{iqx} \epsilon^{abc} \epsilon^{a'b'c'} \langle \mathcal{T} \bigg( (u^T_a  Ch_b)i\gamma^5d_c(x)  (\bar{u}_{a'}  C\bar{h}_{b'}^T) \bar{d}_{c'}(0)i\gamma^5
\nonumber \\
&& -  (u^T_a  Ch_b)d_c(x) (\bar{u}_{a'} C\bar{h}_{b'}^T) \bar{d}_{c'}(0)  \bigg) \rangle  \nonumber \\
& = & \int d^4x  e^{iqx}6  \langle {\rm Tr}  [S_h(x,0) C^T S_u^T(x,0)C] \bigg(  S_d(x,0) - i \gamma_5 C^T S_d^T(x,0) C i \gamma^5  \bigg)  \rangle  \nonumber \\
 & \stackrel{m\rightarrow 0}{\longrightarrow} &
\int d^4x e^{iqx} 12 \pi  \langle {\rm Tr}  \bigg[ S_h(x,0)S^\dagger (0,x)
\bigg] J_{0}(x,0) \rangle,
\label{l-ls-2}
\end{eqnarray}
which is again proportional to $J_0$.  In this case, the current with $h$ replaced by $u$ would be identically zero as the current itself vanishes.

\subsection{Correlation functions of meson with one heavy quark}

Chiral symmetry is also reflected well when there are heavy quarks.
Consider the meson correlation function in Eq.~(\ref{p-s-0}).  If one of the quarks is heavy, the correlation function becomes
\begin{eqnarray}
\Delta_{S-P}^h (q) & = & \int d^4x e^{iqx} \langle \mathcal{T} \bigg( \bar{h} q(x) \bar{q} h(0)- \bar{h}  i\gamma^5 q(x)  \bar{q} i\gamma^5 h(0) \bigg) \rangle  \nonumber \\
& = & \int d^4x  e^{iqx} \langle {\rm Tr}  [ \bigg( i \gamma^5 S(x,0) i \gamma^5 S_h(0,x)-S(x,0) S_h(0,x)   \bigg) ] \rangle  \nonumber \\
 & \stackrel{m\rightarrow 0}{\longrightarrow} &
\int d^4x e^{iqx} \langle {\rm Tr}  \bigg[- ( 0| \frac{1}{D \hspace*{-.2cm}\slash +m_h }|x) J_{0}(x,0)  \bigg( 2 \pi  \bigg)
\bigg] \rangle \label{hp-hs-0}
\end{eqnarray}
Again, the difference vanishes when the current density matrix  of zero modes $J_0$  vanishes as in Eq.~(\ref{BC-0}).

If one further takes the heavy quark limit $m_h \rightarrow \infty$ of Eq.~(\ref{hp-hs-0}). One finds
\begin{eqnarray}
\Delta_{S-P} (q)
 & \stackrel{m=0,m_h \rightarrow \infty}{\longrightarrow} &
\int d^4x e^{iqx} \langle {\rm Tr}  \bigg[- ( 0| \frac{1}{D \hspace*{-.2cm}\slash +m_h }|x) J_{0}(x,0)  \bigg( 2 \pi  \bigg)
\bigg] \rangle \nonumber \\
& = & \frac{2}{m_h} \langle \bar{q} q \rangle.
\label{hp-hs-1}
\end{eqnarray}

One can obtain Weinberg type sum rules\cite{Weinberg:1967kj} for the heavy light quark system by saturating the spectral representation of  Eq.~(\ref{hp-hs-0}) with the ground states of the corresponding channels and then equating it with  Eq.~(\ref{hp-hs-1}).  Then, assuming that the contributions from the excited states and/or continuum cancel, one finds
\begin{eqnarray}
f_{D_0}^2 m_{D_0}^2- f_D^2m_D^2 = 2 m_h \langle \bar{q} q \rangle,
\label{hp-hs-2}
\end{eqnarray}
where the $f_D, f_{D_0}$ and $m_D,m_{D_0}$ are the decay constants and masses of the pseudoscalar and scalar heavy light meson.
The expected scaling of the decay constants and masses with the heavy quark mass are\cite{Aoki:1996xe}
\begin{eqnarray}
f_D & \propto & m_h^{-1/2}, \nonumber \\
m_D & \propto & m_h.
\end{eqnarray}
As expected, Eq.~(\ref{hp-hs-2}) is consistent with the heavy quark scaling, as the light quark condensate on the right hand side does not scale with the heavy quark mass.
More sum rules involving higher moments can be obtained by expanding the  last line of  Eq.~(\ref{hp-hs-0}) with respect to the heavy quark mass.

We can use Eq.~(\ref{hp-hs-2}) for the charm sector to estimate $f_{D_0}$.  Using  $f_D=206.7$ MeV,  $m_D=1865$ MeV, $m_{D_0}=2318$ MeV\cite{pdg}, $m_c=1.26$ GeV and  $\langle \bar{q}q \rangle=(0.225 {\rm GeV})^3 $ used in QCD sum rules\cite{SVZ}, one finds that
\begin{eqnarray}
f_{D_0} =  182 ~{\rm MeV}.
\end{eqnarray}
This is slightly smaller than $f_D$,  as was found to be so in a QCD sum rule calculation\cite{Colangelo:1991ug}.  Lattice calculation also seems to point to a slightly lower value for $f_{D_0}$\cite{Follana:2007uv,Herdoiza:2006qv}.
Similar sum rule can be obtained for the vector and axial vector meson and their corresponding decay constants.
The generalization to medium of the formula in Eq.~(\ref{hp-hs-2}) will tend to modify the masses and the decay constants when chiral symmetry is reduced\cite{Kim:2001xu}.

\subsection{Correlation functions of Baryon with one heavy quark}

In the heavy quark limit $m_h \rightarrow \infty$, Eq.~(\ref{l-ls}) simplifies to the following:
\begin{eqnarray}
\Delta_{\Lambda-\Lambda*}^{m=0} (q)
 & \stackrel{m_h \rightarrow \infty}{\longrightarrow} &
\frac{12 \pi}{m_h}  \langle {\rm Tr}  \bigg[ S^\dagger (0,0)  J_{0}(0,0)
\bigg] \rangle.
\label{C-bh-1}
\end{eqnarray}
Noting that
\begin{eqnarray}
{\rm Re} S^\dagger(0,0) = {\rm Re} S(0,0) = \pi J_0 (0,0),
\end{eqnarray}
the real part of Eq.~(\ref{C-bh-1}) simplifies to the following:
\begin{eqnarray}
\Delta_{\Lambda-\Lambda*}^{m=0} (q)
 & \stackrel{m_h \rightarrow \infty}{\longrightarrow} &
\frac{12 \pi^2}{m_h}  \langle {\rm Tr}  \bigg[ J_0(0,0)  J_{0}(0,0)
\bigg] \rangle.
\label{l-ls-3}
\end{eqnarray}
It should be noted that the operator form of Eq.~(\ref{l-ls-3}) can be obtained directly from the first line of Eq.~(\ref{l-ls}).  Equating this one finds
\begin{eqnarray}
12 \pi^2 \langle {\rm Tr}  \bigg[ J_0(0,0)  J_{0}(0,0)
\bigg] \rangle
&=&\epsilon^{abc} \epsilon^{a'b'c} \langle  (u^T_a i\gamma^5 Cd_b)  (\bar{u}_{a'} i\gamma^5 C\bar{d}_{b'}^T)
\nonumber \\  &&-  (u^T_a  Cd_c) (\bar{u}_{a'} C\bar{d}_{b'}^T)  \rangle
\label{l-ls-4}
\end{eqnarray}

\subsection{4-quark operators}

One also notes from a direct calculation that left hand side of Eq.~(\ref{l-ls-4}) is equal  to a four quark operator composed of quark-antiquark pairs as given in the following identity. \begin{eqnarray}
\epsilon^{abc} \epsilon^{a'b'c} \langle  (u^T_a i\gamma^5 Cd_b)  (\bar{u}_{a'} i\gamma^5 C\bar{d}_{b'}^T)
-  (u^T_a  Cd_c) (\bar{u}_{a'} C\bar{d}_{b'}^T)  \rangle \nonumber \\
= 2 \langle (\bar{u}_a i \gamma^5 d_a)  (\bar{d}_{b} i \gamma^5 u_b) -(\bar{u}_a d_a)  (\bar{d}_{b} u_b)
  \rangle
\label{l-ls-5}
\end{eqnarray}
Hence, one can obtain identity constraints among chiral operators.

The identity of the four quark operators appearing in Eq.~(\ref{l-ls-5}) follows from the fact that both are the same chiral order parameters.
There are other four quark operators that are related to the chiral order parameters.  For example, the four quark operator appearing in the vector and axial vector meson sum rule can be combined into a four quark operator that is also a chiral order parameter\cite{Hatsuda:1991ez}.
\begin{eqnarray}
\langle (\bar{q} \Gamma_1 \lambda^a \tau^b q)  (\bar{q} \Gamma_2 \lambda^a \tau^b q) - (\bar{q} \Gamma_1 i \gamma^5 \lambda^a \tau^b q)  (\bar{q} i \gamma^5 \Gamma_2 \lambda^a \tau^b q)
  \rangle     \nonumber \\
= - \langle {\rm Tr} \bigg[\lambda^2 \tau^2 \Gamma_2 ( 0| \frac{1}{D \hspace*{-.2cm}\slash +m }|0) \Gamma_1 \bigg(
( 0|  \frac{1}{D \hspace*{-.2cm}\slash +m }|0) -i \gamma^5 ( 0| \frac{1}{D \hspace*{-.2cm}\slash +m }|0)  i \gamma^5
  \bigg) \bigg]  \rangle    \nonumber \\
\stackrel{m\rightarrow 0}{\longrightarrow} - \langle {\rm Tr} \bigg[\lambda^2 \tau^2 \Gamma_2 S(0,0) \Gamma_1  J_0(0,0) \bigg( 2 \pi
  \bigg) \bigg]  \rangle.  \label{four-quark-1}
\end{eqnarray}
Here, $\lambda^a$ can either be  the color matrix and/or the unit matrix.  $\tau^a$ is the flavor matrix.  The inclusion of the flavor matrix is important to exclude the disconnected diagram.  It should be noted that if the color and flavor matrix are normalized as $\lambda^2=1$, then the matrix element is independent of whether we include the color matrix or take it to be just the unit matrix.  Also, $\Gamma_1, \Gamma_2$ can be any gamma matrix with either identical and/or different Lorentz index.
It should be noted that Eq.~(\ref{four-quark-1}) can also be written in terms of the left and right handed quarks as follows:
\begin{eqnarray}
2\langle (\bar{q} \Gamma_1 \lambda^a \tau^b q_R)  (\bar{q}_L \Gamma_2 \lambda^a \tau^b q) & +& (\bar{q} \Gamma_1 \lambda^a \tau^b q_L)  (\bar{q}_R \Gamma_2 \lambda^a \tau^b q) \rangle.  \label{four-quark-2}
\end{eqnarray}

Several identities follow from Eq.~(\ref{four-quark-1}) or Eq.~(\ref{four-quark-2}).  For example,
\begin{eqnarray}
\langle (\bar{u}  \lambda^a d)  (\bar{d}  \lambda^a u)  -(\bar{u}  i \gamma^5 \lambda^a d)  (\bar{d} i \gamma^5 \lambda^a u)
  \rangle    & = &
\langle (\bar{u}  d)  (\bar{d}  u)  - (\bar{u}  i \gamma^5 d)  (\bar{d} i \gamma^5  u)
  \rangle  \nonumber \\
   & = &
2 \langle (\bar{u}_L  d_R)(\bar{d}_L u_R) + (\bar{d}_R  u_L)(\bar{u}_R d_L)
  \rangle, \nonumber \\
\end{eqnarray}
which vanishes when chiral symmetry is restored.
One can take $\Gamma_1$ and $\Gamma_2$ to have Lorentz indices and derive identities involving twist-4 operators\cite{Jaffe:1981td,Jaffe:1982pm,Ellis1,Ellis2,Choi:1993cu,Lee:1993ww}.

\section{$U_A(1)$ effect and Topological configurations}

The integration over gauge fields given in Eq.(\ref{measure0}) can have nontrivial configurations characterized by $\nu=(g^2/32 \pi^2) \int d^4x F \tilde{F} \neq 0$.  The QCD partition function can be divided and summed over by different topological configurations.
\begin{eqnarray}
Z= \sum_{\nu} Z_\nu.
\end{eqnarray}
Configuration with nontrivial topological charges  have $n_+$ ($n_-$) number of right handed (left handed) zero mode solutions that satisfy the index theorem $\nu=n_+-n_-$.  For these configurations it is useful to explicitly write the zero mode contribution with asymmetric chirality  as a separate quark determinant.  For example, for configurations with $\nu=1$, the partition function is given as follows:
\begin{eqnarray}
Z_{\nu=1} & = & \int [dA_\mu(x)]_{\nu=1}
[d \bar{q}(x) d q(x)]' [d \bar{q}_L d q_R]  \exp \bigg[-S_{YM} -\int d^4x \bar{q}(x)[D \hspace*{-.2cm}\slash +m]q(x) \bigg] \nonumber \\
& = &
\int [dA_\mu(x)]_{\nu=1}  \exp [-S_{YM}]
{\rm det}' [D \hspace*{-.2cm}\slash +m] \times {\rm det} \big( \int d^4x \bar{\psi}_0(x) m \psi_0(x) \bigg)  \nonumber \\
& = & \int [d \mu]_{\nu=1}{\rm det} \big( \int d^4x \bar{\psi}_0(x) m \psi_0(x) \bigg). \label{measure1}
\end{eqnarray}
Here,   $\psi_0$ is the zero mode solution and the measure in the zero mode runs over all flavors. The prime after the quark measure  means that the zero mode is taken out; the integration over these quarks give the primed quark determinant.  The second determinant  comes from the integral over the zero modes.  One should note that the contribution from these configurations is zero in the chiral limit.  Therefore, the path integral for the difference in the correlation functions discussed in the previous section can be thought of the integration  over gauge field configurations with $\nu=0$.

However when calculating the $n$-point functions involving quark operators with asymmetric chirality containing all flavors, the topological configuration will give non vanishing contributions.  This is so because in such $U_A(1)$ variant configurations, the integral over the chiral zero modes are saturated by the external operators and do not appear in the zero mode part of the quark determinant.

Let us consider the 2 flavor (u,d quarks) case.  For the quark condensate,
\begin{eqnarray}
\langle \bar{q}q(0) \rangle & = & \langle \bar{q}q(0) \rangle_{\nu=0} + \frac{1}{Z}
\int [dA_\mu(x)]_{\nu=1}
[d \bar{q}(x) d q(x)]' \nonumber \\
 && \times [d \bar{u}_L d u_R] [d \bar{d}_L d d_R]  \bigg( \bar{q}_Lq_R(0) \bigg) \exp \bigg[-S_{YM} -\int d^4x \bar{q}(x)[D \hspace*{-.2cm}\slash +m]q(x) \bigg] \nonumber \\
&=& \langle \bar{q}q(0) \rangle_{\nu=0} + \frac{-1}{Z} \int [d \mu]_{\nu=1} 2 \bigg( \bar{\psi}_0 \psi_0(0)  \bigg) \bigg( \int d^4x \bar{\psi}_0(x) m\psi_0(x) \bigg) + \cdot .
\end{eqnarray}
As can be seen in the second term of the above equation, the topological configurations are proportional to the light quark mass and do not contribute to the quark condensate.
However, when one looks at the two point function, the zero mode integral can be saturated by the external operators:
\begin{eqnarray}
\langle \bar{q}q(x) \bar{q}q(0) \rangle & = & \langle \bar{q}q(x) \bar{q}q(0) \rangle_{\nu=0} \nonumber \\
&& + \frac{1}{Z} \int [d \mu]_{\nu=1} [du_R][d\bar{u}_L] [dd_R][d\bar{d}_L] \bigg( \bar{q}_Lq_R(x)  \bigg) \bigg( \bar{q}_Lq_R(0) \bigg) + \cdot \cdot \nonumber \\
&=& \cdot \cdot + \frac{1}{Z} \int [d \mu]_{\nu=1} 2 \bigg( \bar{\psi}_0 \psi_0(x)  \bigg) \bigg( \bar{\psi}_0 \psi_0(0) \bigg) + \cdot \cdot .
\end{eqnarray}
The topological configurations will therefore contribute when looking at the difference between correlation functions related by the $U_A(1)$ transformation.
For example,
\begin{eqnarray}
\langle \bar{q}q(x) \bar{q}q(0) \rangle & -& \langle \bar{q} \tau^3 q(x) \bar{q} \tau^3 q(0) \rangle \nonumber \\
   && =  \langle .. \rangle_{\nu=0}  + \frac{1}{Z} \int [d \mu]_{\nu=1} 4 \bigg( \bar{\psi}_0 \psi_0(x)  \bigg) \bigg( \bar{\psi}_0 \psi_0(0) \bigg) + \cdot \cdot .
\end{eqnarray}

For the case of $N_f$ flavors, $\nu=1$ configuration will still be proportional to  $m^{N_f-2}$ and  not contribute to the four quark operators.  In this case, the lowest dimensional operators where the $\nu=1$ configuration contribute are the $2N_f$ quark operators\cite{Lee:1996zy}.

Summarizing the sections so far, one can conclude that chiral symmetry breaking occurs when the current density matrix of zero modes are non zero, while the $U_A(1)$ breaking effect comes about when the topological configuration contributes to the operator expectation value  and thus in principle independent of chiral symmetry breaking.  Phenomenologically, it means that even if chiral symmetry is restored at finite temperature or density, the $U_A(1)$ breaking effect will persist in $n$-point functions and will lead to phenomenologically  observable consequences.  However, as discussed before, how big "$n$" has to be depends on the number of flavor, and how exact the statements are depends of the magnitude of explicit symmetry breaking.  One of the obvious physical observable is the mass of  a hadron.  Therefore, we will look at how the relations on the $U_A(1)$ effect is related to the $\eta'$ mass.

\section{Gluonic correlation}

\subsection{LET for Gluonic operators}

Let us first review the derivation of the low energy theorems for gluonic correlations\cite{Lee:2000cdb}.  We define the scalar and pseudo-scalar gluonic two-point function,
\begin{eqnarray}
S(q) & = & i \int d^4x \,e^{iq x} \left\langle \mathcal{T}\frac{ 3 \alpha_s}{4 \pi} G^2(x)  \frac{ 3 \alpha_s}{4 \pi} G^2(0) \right\rangle,  \nonumber \\
P(q) & = & i \int d^4x \,e^{iq x}\left\langle \mathcal{T}\frac{ 3 \alpha_s}{4 \pi} G  \tilde{G}(x)  \frac{ 3 \alpha_s}{4 \pi} G \tilde{G}(0) \right\rangle.
	\label{sg-pg}
\end{eqnarray}
Here, $G^2=G^a_{\mu \nu} G^a_{\mu \nu}$ and $G \tilde{G}=G^a_{\mu \nu} \frac{1}{2} \epsilon_{\mu \nu \alpha \beta} G^a_{\alpha \beta}$.
There is a well known low energy theorem that follows from noting that the measure given in Eq.~(\ref{measure0}) leads to the following identity:
\begin{eqnarray}
\frac{d}{d(-1/4g^2)} \langle O \rangle
= i \int d^4x \langle \mathcal{T} O(x) g^2G^2(0) \rangle, \label{der-g}
\end{eqnarray}
where the bare coupling $g^2$ is related to the ultraviolet cutoff  $M_0$ to the unique scale of QCD.
\begin{equation}
	\Lambda=M_0 \exp \left( -\frac{8 \pi^2}{bg_s^2} \right),
	\label{eq12}
\end{equation}
where $b=11-\frac{2}{3}N_f$ \cite{NSVZ81}.  If there are no other scales in the system, the physically observable matrix element should be proportional to the $d$'th power of the scale, where $d$ is the dimension of the operator.
Therefore,
\begin{eqnarray}
\frac{d}{d(-1/4g^2)} \langle O \rangle & = &
\frac{d}{d(-1/4g^2)} \bigg( {\rm const} \times \Lambda^d  \bigg) \nonumber  \\
&= & d\frac{32 \pi^2}{b} \langle O \rangle  \label{scaling}
\end{eqnarray}
Substituting $O=(3 \alpha_s /4 \pi)G^2$ to the left hand side of the first line of Eq.~(\ref{sg-pg}) leads to the following low energy theorem for the scalar correlator.
\begin{eqnarray}
\frac{3}{16 \pi^2} \frac{d}{d(-1/4g^2)} \langle \frac{3 \alpha_s }{4 \pi}G^2 \rangle & = &
i \int d^4x \langle \mathcal{T} \frac{3 \alpha_s }{4 \pi}G^2(x) \frac{3 \alpha_s }{4 \pi} G^2(0) \rangle,
\end{eqnarray}

The situation for the pseudo-scalar is more involved.
The low energy theorem depends on whether the correlation function is  calculated in pure gauge theory or in the presence of light quarks.  When there are light quarks present, the pseudo-scalar gluon current can be written as
\begin{eqnarray}
\frac{ N_f \alpha_s}{4 \pi} G \tilde{G} =\sum_q \partial_\mu \bar{q} \gamma_\mu \gamma_5 q. \label{axial}
\end{eqnarray}
Therefore, for $N_f=3$, $P(q)$ can be written in terms of the quark currents.
\begin{align}
P(q)  =&  q^\mu q^\nu i\int d^4x\,e^{iq\cdot x}
\langle\mathcal{T} \bar{q} \gamma_\mu \gamma_5 q(x) \,
\bar{q} \gamma_\nu \gamma_5 q(0) \rangle.
\label{ps-1}
\end{align}
Hence,
\begin{eqnarray}
P(q=0) =0 \label{ps-full-0}
\end{eqnarray}
However, when there are no light quarks, there are similar low energy theorems as in the scalar gluonic correlator.
To derive a similar equation for the pseudo-scalar current, first introduce a $\theta$ term as in Eq.~(\ref{z-theta}).  Then we can have expectation value of parity odd operators.   In such a theory,
\begin{eqnarray}
\frac{d}{d(-1/4g^2)} \langle \frac{3 \alpha_s }{4 \pi}G \tilde{G} \rangle_\theta & = &
i \int d^4x \langle \mathcal{T} \frac{3 \alpha_s }{4 \pi}G \tilde{G}(x) g^2G^2(0) \rangle_\theta,
\end{eqnarray}
Using the heavy quark mass expansion, we can write it into the following form,
\begin{eqnarray}
\frac{d}{d(-1/4g^2)} \langle \frac{3  }{4 }(-8) m_h \bar{h} i \gamma_5 h  \rangle_\theta & = &
i \int d^4x \langle \mathcal{T} \frac{3 \alpha_s }{4 \pi}G \tilde{G}(x) (4 \pi^2) (-12) m_h \bar{h} h (0) \rangle_\theta. \label{ps-2}
\end{eqnarray}
As explained in Appendix C, we make a chiral rotation of the heavy quark by $\pi/2$: This will only change the heavy quark mass matrix to $m_h \rightarrow  m_h i\gamma_5$.  Then Eq.~(\ref{ps-2}) will change to the following:
\begin{eqnarray}
\frac{d}{d(-1/4g^2)} \langle \frac{3  }{4 }(8) m_h \bar{h}  h  \rangle_\theta & = &
i \int d^4x \langle \mathcal{T} \frac{3 \alpha_s }{4 \pi}G \tilde{G}(x) (4 \pi^2) (-12) m_h \bar{h} i \gamma_5 h (0) \rangle_\theta.
\end{eqnarray}
Making use of the heavy quark mass expansion again, we obtain the following relation for the low energy theorem for the pseudo scalar current.
\begin{eqnarray}
-\frac{1 }{12 \pi^2 }\frac{d}{d(-1/4g^2)} \langle   \frac{3 \alpha_s }{4 \pi}G^2  \rangle & = &
i \int d^4x \langle \mathcal{T} \frac{3 \alpha_s }{4 \pi}G \tilde{G}(x) \frac{3\alpha_s }{4\pi}G \tilde{G} (0) \rangle,
\end{eqnarray}
where we have taken the $\theta$ to be zero  at the end.
Although we have used the heavy quark mass expansion, this result should be valid as we can take the mass to be infinitely heavy.
Using the formula in Eq.~(\ref{scaling}), we finally obtain the low energy theorem for both the scalar and pseudo scalar gluonic correlators.
\begin{eqnarray}
S(q=0) & = &\frac{18}{b} \langle \frac{\alpha_s}{\pi}G^2 \rangle \nonumber \\
P_0(q=0) & = & - \frac{8}{b} \langle \frac{\alpha_s}{\pi}G^2 \rangle,
\label{LET-1}
\end{eqnarray}
where the subscript $0$ in the pseudo-scalar current represents calculation in pure gauge theory.

At finite temperature, the constant gets modified because there are extra scales in the system; temperature and/or density.  Then the matrix element could depend directly on the extra scale.  Hence, one has to make sure that the derivative in Eq.~(\ref{der-g}) do not act on these part.  Such operations can be taken into account explicitly by noting that the matrix element can now be expressed as follows:
\begin{eqnarray}
\langle O \rangle_{\mu, T}  & = & {\rm const} \times \Lambda^d f \bigg( \frac{T}{\Lambda}, \frac{\mu}{\Lambda} \bigg),
\end{eqnarray}
where the subscript $\mu,T$ denote that the expectation value is taken  at finite chemical potential and/or temperature.  Eq.~(\ref{der-g}) will then be modified as
\begin{eqnarray}
\frac{d}{d(-1/4g^2)} \langle O \rangle_{\mu, T}
&= & \frac{32 \pi^2}{b} \bigg(d-T\frac{\partial}{\partial T}-\mu \frac{\partial}{\partial \mu} \bigg) \langle O \rangle,  \label{scaling-2}
\end{eqnarray}
and the low energy theorem will look as follows:
\begin{eqnarray}
S(q=0) & = &\frac{9}{2b} \bigg(d-T\frac{\partial}{\partial T}-\mu \frac{\partial}{\partial \mu} \bigg) \langle \frac{\alpha_s}{\pi}G^2 \rangle \nonumber \\
P_0(q=0) & = & - \frac{2}{b}\bigg(d-T\frac{\partial}{\partial T}-\mu \frac{\partial}{\partial \mu} \bigg) \langle \frac{\alpha_s}{\pi}G^2 \rangle
\label{LET-2}
\end{eqnarray}

\subsection{Gluonic correlator in medium}

Now let us go back to the pseudo scalar correlation function given in Eq.~(\ref{sg-pg}) and discuss their fate when chiral symmetry is restored.
Here, we will again introduce light quarks with $N_f$ flavors.  Then using Eq.~(\ref{axial}), the pseudoscalar gluonic current can be written in terms of the divergence of the axial current.
\begin{align}
P(q)  =&  q^\mu q^\nu i\int d^4x\,e^{iq\cdot x}
 \bigg[\langle\mathcal{T} \bar{q} \gamma_\mu \gamma_5 q(x) \,
\bar{q} \gamma_\nu \gamma_5 q(0) \rangle
 -
\langle\mathcal{T} \bar{q}  \gamma_\mu q(x) \,
\bar{q}  \gamma_\nu  q(0) \rangle \bigg], \label{coupling}
\end{align}
where we have subtracted out the zero contribution from the conserved vector current.  Using the previous representations, the correlation function can be written as
\begin{align}
P(q) =&  q^\mu q^\nu \int d^4x e^{iq \cdot x}  \bigg[ \langle    {\rm Tr} [S(x,x)  \gamma_\mu \gamma_5]
 {\rm Tr} [S(0,0) \gamma_\nu \gamma_5 ]
-  {\rm Tr} [S(x,x)  \gamma_\mu ]  {\rm Tr} [S(0,0) \gamma_\nu  ]   \rangle \nonumber \\
&- \langle    {\rm Tr} [S(0,x)  \gamma_\mu \gamma_5 S(x,0) \gamma_\nu \gamma_5 ]
+  {\rm Tr} [S(0,x)  \gamma_\mu S(x,0) \gamma_\nu  ]   \rangle
 \bigg].
\label{sumrule}
\end{align}
Now, when chiral symmetry is restored, the two terms in the second line of the above equation will cancel each other, as they are the same as the difference between flavored chiral partners; vector and axial vector currents.   The remaining first line constitutes the disconnected contributions.
However, as discussed before and in appendix Eq.~(\ref{disconnected-1}), the disconnected pieces all vanish in the chiral limit when chiral symmetry is restored.  This is again because the disconnected contributions are proportional to the current density matrix of zero eigenvalues.
\begin{eqnarray}
{\rm Tr} [S_A(x,x)] \sim {\rm Tr} [S_A(x,x) \Gamma ] = {\rm Tr} [J_{\lambda=0}(x,x) \Gamma]  \rightarrow 0,
\label{SA}
\end{eqnarray}
where $\Gamma$ is a Hermitian gamma matrix\cite{Cohen:1996ng}.
Therefore, when chiral symmetry is restored,
\begin{eqnarray}
P(q) \rightarrow 0, \label{ps-chiral}
\end{eqnarray}
in the chiral limit for any finite external momenta $q$.  It should be noted that when Eq.~(\ref{ps-chiral}) is true only when chiral symmetry is restored, while Eq.~(\ref{ps-full-0}) is always true.

\section{$\eta'$ mass in medium}

We have discussed the relationship between chiral symmetry and $U_A(1)$ effects in both the vacuum and in medium.  When chiral symmetry is restored, the correlation functions between chiral partners will become degenerate.  In addition, the two point correlation functions that are related by $U_A(1)$ and chiral transformations will become degenerate when the number of massless flavors is larger than two.  For the physical case, this means that the effect of $U_A(1)$ symmetry in the two point functions will only be proportional to the strange quark mass.   Here, we will review how such relation can be related to the fate of the  $\eta'$ mass when chiral symmetry is restored\cite{Kwon:2012vb}.

\subsection{Witten-Veneziano formula}

One can directly relate the pseudo-scalar current and the $\eta'$ mass using the large $N_c$ arguments\cite{Witten:1979vv,Veneziano:1979ec}.

In terms of the physical states, the pseudo-scalar gluonic correlation function looks as follows:
\begin{eqnarray}
P(q) & = &-\sum_{n}\frac{|\langle0|\frac{3 \alpha}{4 \pi} G\tilde{G}|\,\text{$n^\textrm{th}$ glueball}\rangle|^2}{q^2-M_n^2} \nonumber \\
		&& -\sum_{n}\frac{|\langle0|\frac{3 \alpha}{4 \pi}G\tilde{G}|\,\text{$n^\textrm{th}$ meson}\rangle|^2}{q^2-m_n^2} \nonumber \\
		  &\equiv & P_0(q)+P_1(q).
	\label{eq2}
\end{eqnarray}
The first term in the right hand side of the above equation indicates contributions from glueballs,  while the second term shows those from the mesons composed of light quarks.
As discussed before, $P(q=0)=0$ while $P_0(q=0) \neq 0$ when we assume massless quark.  This seems in contradiction to the large $N_c$ argument because the $P_0(q=0)$ that scales as $N_c^2$ is canceled by quark effects that scales as order $N_c$.  It was noted that this cancelation is possible by the existence of the $\eta'$ whose mass scales as $1/N_c$ and cancels the gluonic effect in Eq.\,(\ref{eq2}) only when $q=0$; other meson masses have a smooth large $N_c$ limit.  This constraint directly relates the $\eta'$ mass to the low energy theorem.
\begin{eqnarray}
P_0(0)=-\frac{|\langle0|\frac{3 \alpha}{4 \pi} G\tilde{G}|\eta^\prime\rangle|^2}{m_{\eta^\prime}^2}.
	\label{eq4}
\end{eqnarray}
By using the $U_A(1)$ anomaly,
\begin{equation}
\begin{split}
	 \langle0|\frac{3 \alpha}{4 \pi} G\tilde{G}|\eta^\prime\rangle&=\langle0|\partial_\mu J_5^\mu|\eta^\prime\rangle\\
					 &=\sqrt{N_f}\,m_{\eta^\prime}^2f_\pi,
\end{split}
	\label{eq5}
\end{equation}
Eq.\,(\ref{eq4}) becomes as follows:
\begin{equation} P_0(0)=-m_{\eta^\prime}^2f_\pi^2 N_f,
	\label{eq6}
\end{equation}
where $N_f$ is the number of light flavors. In Eq.\,(\ref{eq5}), we made use of $f_{\eta^\prime}=f_\pi$ to lowest order in $N_c$.  Eq.\,(\ref{eq6}) is the celebrated WV formula.

Substituting Eq.~(\ref{LET-1}) into Eq.~(\ref{eq6}), we find
\begin{eqnarray}
m_{\eta^\prime} =\sqrt{\frac{8}{33}} \frac{1}{f_\pi} \langle \frac{\alpha_s}{\pi}G^2 \rangle^{1/2} \approx 464 ~~{\rm MeV},
\label{vacuum-mass}
\end{eqnarray}
where we have used $f_\pi=130\,\mathrm{MeV}$, $ \langle \frac{\alpha_s}{\pi}G^2 \rangle= (0.35 {\rm GeV})^4$ and used $b=11$ for pure glue theory.
This is smaller than the vacuum value of the $\eta^\prime$ mass as expected; this is the part coming from the $\mathrm{U}_A(1)$ effect to the  mass of $\eta^\prime$.

\subsection{Witten-Veneziano formula in medium}

To obtain the generalized formula at finite temperature, one notes that the thermal gluonic effects are of order $N_c^2$ while that of the quarks are of order $N_c$.  If one is in the confined phase, the phase is composed of mesons, glueballs and baryons: The scattering of these states scale as order 1, 1 and $N_c$ respectively.  Hence as long as one assumes that the number of hadrons do not scale with $N_c$, hadronic effects can be neglected.  Near the phase transition, the degeneracy of hadrons would increase and follow the scaling of gluons and quarks.  Therefore, the leading order effect would come from the gluons.  Taking into account the effect of the thermal gluons, the low energy theorem given in Eq.~(\ref{LET-2}) will be modified only by the effect of finite temperature.  Therefore,  the generalization of the WV formula given in Eq.~(\ref{eq4}) would be the following.
\begin{equation}
 m_{\eta^\prime}^2 =
  \frac{|\langle 0|\frac{3\alpha_s}{4\pi}
  G\tilde{G}|\eta^\prime \rangle |^2}{\frac{2}{b}\left(
						  d-T\frac{\partial}{\partial
						  T} \right)
  \left\langle \frac{\alpha_s}{\pi}G^2 \right\rangle_{T,\text{pure
  gauge}}}.
  \label{eq16}
\end{equation}
In obtaining the result, we have assumed that the general structure given in Eq.~(\ref{eq2}) does not change.  Naively, finite temperature effect will introduce the medium rest frame and break Lorentz invariance; this means that the correlation function is a separate function of $q_0$ and $\vec{q}$.  However, the thermal correction we consider is embedded in the scalar gluon condensate and in the scalar low energy theorem. Therefore, the corresponding change should also be only in the scalar quantity; hence Eq.~(\ref{eq2}) should be valid for our purpose.

First let us consider the denominator of Eq.\,(\ref{eq16}).
It has been known for a long time, that the gluon condensate has
contribution from the perturbative and non-perturbative parts.
Moreover, it was also known that at the critical temperature, the
non-perturbative contribution changes abruptly, but does not vanish
completely, and retains more than half of its non-perturbative
value~\cite{Lee:1989qj,Adami:1990sv,Brown:2006bp}.


The effect of subtracting out the second term in the denominator of Eq.\,(\ref{eq16}) is to get rid of the perturbative
correction, or the seemingly scale breaking effect that is not related
to scale breaking but due to the introduction of an external scale
parameter $T$.  The leading perturbative correction to the gluon
condensate is proportional to
$g_s^4T^4$~\cite{Kapusta:1979fh,Boyd:1996bx}.  Therefore, assuming that
the temperature dependence is of the following form,
\begin{equation}
	\left\langle \frac{\alpha_s}{\pi} G^2 \right\rangle_T=G_0(T) +ag_s^4T^4,
	\label{eq17}
\end{equation}
we find,
\begin{equation}
	\left(d - T \frac{\partial }{\partial T}\right) \left\langle \frac{\alpha_s}{\pi} G^2 \right\rangle_T =\left(d - T \frac{\partial }{\partial T}\right) G_0(T),
	\label{eq18}
\end{equation}
if the temperature dependence  of $g_s$ is neglected.
The only temperature dependence that survives is $G_0(T)$, whose scale
dependence is coming from dimensional transmutation and not from the
external temperature only.
It is the non-perturbative part that dominates the behavior of the right
hand side of Eq.\,(\ref{eq16}).  Moreover, as we discussed before, for the gluon condensate,  one has to use the lattice result obtained in the pure-gauge theory calculation.  The critical temperature in such a calculation $T_{pure-gauge}\sim$ 260 MeV is known to be around 100 MeV larger than that from a full QCD calculation $T_{QCD}$\cite{Boyd:1996bx}.  On the other hand, while the expected change of the gluon condensate is more abrupt in the pure-gauge calculation, the actual change in the condensate value itself at the critical temperature is found to be similar to the full calculation\cite{Morita:2007hv}.
This means that the change of the gluon condensate can be
effectively neglected up to temperatures near $T_{QCD}$.

Finally, we use the fact that when chiral symmetry is restored, Eq.~(\ref{eq16}) is zero for large $q$ as given by  Eq.~(\ref{ps-chiral}).
Hence, when chiral symmetry is restored,
\begin{eqnarray}
\langle 0|G\tilde{G} | \eta^\prime \rangle \sim O(m_q).
\end{eqnarray}
Therefore, going back to
Eq.\,(\ref{eq16}) and making use of the previous discussions,
we find that when chiral symmetry is restored,
\begin{eqnarray}
m_{\eta^\prime}^2
\stackrel{\langle \bar{q}q \rangle \rightarrow 0}{\longrightarrow}  0 , \label{final}
\end{eqnarray}
in the chiral limit.
One concludes that in the large $N_c$ limit of QCD, $\eta^\prime$
mass will become degenerate with the other Goldstone bosons.
A similar conclusion was obtained in Ref.~\cite{Benic:2011fv}; that
the anomalous $\mathrm{U}_A(1)$ $\eta^\prime$ mass squared vanishes at high T as the chiral quark condensate  $\langle \bar{q}q \rangle  $.

\section{Conclusions}

We have used Euclidean path integral to show that the current density matrix of zero modes is responsible for the breaking of chiral symmetry in operators constructed from correlation functions.  This is a generalization of the Banks-Casher formula to any order parameter.   As for the  $U_A(1)$ symmetry, the contribution of  topologically non trivial configurations determine  the presence of its breaking effect.   The contribution of topological configurations however depend on the number of flavors and the number of external legs of the correlator.  We have  also discussed how the mechanism can be applied to the gluon correlation functions.  Combining the results, we have described how the Witten Veneziano formula for the $\eta'$ mass can be generalized to the medium.

\section*{Acknowledgements}

This work was supported by Korea national research foundation under grant number KRF-2011-0030621 and KRF-2011-0015467.

\appendix

\section{Notations}

Here we will summarize the notations.  The QCD partition function is given by
\begin{eqnarray}
Z= \int [dA_\mu(x)][dq(x) d\bar{q}(x)] \exp [iS_{QCD}],
\end{eqnarray}
where
\begin{eqnarray}
S_{QCD}=\int d^4x \bigg[ -\frac{1}{4g^2} F^2
+\bar{q}( i \gamma^\mu(\partial_\mu -iA_\mu)-m_q) q \bigg].
\end{eqnarray}

The Wick rotation to imaginary time,
\begin{eqnarray}
x_0=t \rightarrow -i\tau=-ix_4
\end{eqnarray}
comes with the following:
\begin{eqnarray}
d^4x & \rightarrow & -id^4 x_E  \\
A_0 & \rightarrow & i A_4 \\
F_{\mu \nu} F^{\mu \nu} =-F_{0i}^2+F_{ij}^2 & \rightarrow & F_{4i}^2 +F_{ij}^2 \\
i \gamma^\mu (\partial_\mu -iA_\mu) & \rightarrow & i \gamma_0 (\partial_0-iA_0)-i \gamma_i (\partial_i -iA_i) \nonumber \\
& \rightarrow & - \gamma_0 (\partial_4-iA_4)-i \gamma_i (\partial_i -iA_i) = -\gamma_\mu^E( \partial_\mu-iA_\mu),
\end{eqnarray}
where $\gamma_4^E=\gamma_0, \gamma_i^E=i \gamma_i$.
It should be noted that the gauge field here is redefined such that the field strength tensor is independent of the coupling.  One can go back to the usual definition of the gauge field ${\cal A}_\mu$ and field strength tensor $G_{\mu \nu}$ by redefining the following substitution.
\begin{eqnarray}
A_\mu & = & g {\cal A}_\mu \nonumber \\
\frac{1}{4g^2} F^2 & = & \frac{1}{4} G^2
\end{eqnarray}

The partition function then becomes
\begin{eqnarray}
Z= \int [dA_\mu(x)][dq(x) d\bar{q}(x)] \exp [-S^E_{QCD}], \label{eu-z}
\end{eqnarray}
where
\begin{eqnarray}
S^E_{QCD}=\int d^4x_E \bigg[ \frac{1}{4} G^2
+\bar{q}(  \gamma_\mu^ED_\mu+m_q) q \bigg], \label{eu-l}
\end{eqnarray}
and  $D_\mu=\partial_\mu -igA_\mu$.  Also, $S_{YM}$ will be used for the gluonic part only.

After integrating over the quark field, the partition function in Eq.~(\ref{eu-z}) can be written as,
\begin{eqnarray}
Z & = & \int [dA_\mu(x)] \exp [- S_{YM} ] {\rm det} [D \hspace*{-.2cm}\slash +m]   \nonumber \\
& = & \int [d \mu]  \label{measure01}
\end{eqnarray}
\begin{enumerate}
\item
One should note that $d \mu$ is a positive definite measure.
To prove this, one notes that the eigenvalues of the determinant comes in pairs.  If $\psi$ is an eigenvector $ i D \hspace*{-.2cm}\slash \psi=\lambda \psi$, then so is $\gamma^5 \psi$ because $ i D \hspace*{-.2cm}\slash \gamma^5 \psi=-i \gamma^5 D \hspace*{-.2cm}\slash \psi=-\lambda \gamma^5 \psi$.  Hence,
\begin{eqnarray}
{\rm det} [D \hspace*{-.2cm}\slash +m]=\prod_\lambda (m-i\lambda)
=\prod_{\lambda >0} (m^2+\lambda^2) >0
\end{eqnarray}

\item  Sometimes it is useful to write the determinant part as
\begin{eqnarray}
Z & = & \int [dA_\mu(x)] {\rm det} [D \hspace*{-.2cm}\slash +m]  \exp [- S_{YM}] \nonumber \\
& = & \int [dA_\mu(x)]  e^{  {\rm Tr}[\ln(D \hspace*{-.2cm}\slash +m)]  } \exp [- S_{YM}] \label{trace-log}
\end{eqnarray}

\item The vacuum expectation of an operator ${\cal O}(x)$ is defined as,
\begin{eqnarray}
\langle {\cal O}(x) \rangle & = & \frac{1}{Z} \int [d \mu] {\cal O}(x)
\end{eqnarray}
\end{enumerate}

\subsection{Derivation of Banks-Casher formula}

Here, we repeat the derivation of the Banks-Casher formula~\cite{BC} that is extensively used in numerical simulations to calculate chiral symmetry breaking.
The chiral order parameter is defined as follows:
\begin{eqnarray}
\langle \bar{q} q \rangle & = &  \frac{1}{V} \int d^4 x \langle \bar{q} q(x) \rangle = -\frac{1}{V} \int d^4 x {\rm Tr} [\langle  q(x)\bar{q}(x) \rangle]
\nonumber \\
&=&
-\frac{1}{V} \int d^4 x \langle {\rm Tr} [ ( x| \frac{1}{D \hspace*{-.2cm}\slash +m }|x) ] \rangle
\nonumber \\
&=&
- \langle {\rm Tr} [ ( 0| \frac{1}{D \hspace*{-.2cm}\slash +m }|0) ] \rangle \nonumber \\
&=&
-\frac{1}{Z}
\int [d \mu ] {\rm Tr} [ ( 0| \frac{1}{D \hspace*{-.2cm}\slash +m }|0)],
\end{eqnarray}
where in the first line, the measure given in Eq.~(\ref{eu-z}) has been used.
One notices that the quark propagator selects to divide out the particular matrix element from the quark determinant.

To further calculate the trace of the inverse quark propagator, one can use the solutions of the Dirac equation and express the chiral condensate in terms of the wave functions as follows:
\begin{eqnarray}
\langle \bar{q} q \rangle
&=& -\int d^4x \langle \sum_\lambda  \frac{\psi_\lambda(x)^\dagger \psi_\lambda(x) }{V} \frac{1}{m-i\lambda} \rangle \nonumber\\
&=& -\langle  \sum_\lambda  \rho(\lambda) \frac{1}{m-i\lambda} \rangle  .
\label{BC-1}
\end{eqnarray}
To discuss chiral symmetry breaking in the chiral limit, we have to look at $\lim_{m\rightarrow 0} \langle \bar{q} q \rangle$.  To this end, we note,
$\lim_{m \rightarrow 0} \frac{1}{m-i\lambda} =i {\rm P}(\frac{1}{\lambda})+\pi \delta(\lambda)$.  Now, the imaginary part is zero because the eigenvalues are paired (positive and negative eigenvalues) and the eigenfunction of the pairs  are related by $\gamma^5$ so that $\psi^\dagger_{-\lambda}(x) \psi_{-\lambda}(x)=\psi^\dagger_{\lambda}(x) \psi_{\lambda}(x)$.  Hence,
\begin{eqnarray}
\langle \bar{q}q \rangle=- \pi \langle  \sum_{\lambda=0}  \psi^\dagger_\lambda(0) \psi_\lambda(0) \rangle  =-\pi \langle {\rm Tr} [J_{\lambda=0}(0,0)] ) \rangle.
\end{eqnarray}
where the current density matrix is defined in Eq.~(\ref{current-density-matrix}).
A simple booking way to understand the meaning of $\sum_{\lambda=0}$ is to assume that the original sum of eigenvalues consisted of continuous eigenvalues and a degenerate set of finite zero eigenvalues.  The remaining sum is the sum over finite set of remaining zero eigenvalues.
This relation shows that the  essence of chiral symmetry breaking is the non-vanishing current density matrix of zero eigenvalues.  For a system of finite box of volume $V$, the zero modes are small eigenvalues that vanish in the infinite box limit.  In practice, the wave functions of zero modes are expected to be localized around a topological gauge configuration and hence chiral symmetry breaking can be translated to non vanishing expectation value of the matrix  $ \psi_{0}(0) \psi_{0}^\dagger(0)$.

Another derivation comes from   using  Eq.(\ref{eu-z}) and Eq.(\ref{eu-l}) so that,
\begin{eqnarray}
\langle \bar{q}q \rangle & = &  \frac{1}{V}    \langle \int d^4 x \bar{q}q(x) \rangle \nonumber \\
&= & -\frac{1}{V}   \frac{\partial Z}{Z \partial m_q} \nonumber \\
& = &
-\frac{1}{V} \int d^4 x {\rm Tr} [ \langle x| \frac{1}{D \hspace*{-.2cm}\slash +m }|x \rangle ]
\end{eqnarray}
where the last line follows using Eq.(\ref{trace-log}).

The  disappearance of the imaginary part in Eq.~(\ref{BC-1}) can be  implemented at the operator level  by the following method:
\begin{eqnarray}
\langle \bar{q} q \rangle & = &  -  \langle {\rm Tr} \bigg[ ( 0| \frac{1}{D \hspace*{-.2cm}\slash +m }|0) \bigg] \rangle  \nonumber \\
&=&
- \frac{1}{2} \langle {\rm Tr} \bigg[ ( 0| \frac{1}{D \hspace*{-.2cm}\slash +m }|0) -i \gamma^5 ( 0| \frac{1}{D \hspace*{-.2cm}\slash +m }|0)  i \gamma^5 \bigg] \rangle \nonumber \\
& \stackrel{m\rightarrow 0}{\longrightarrow} &
- \frac{1}{2} \langle {\rm Tr} \bigg[ \sum_{\lambda} \psi_{\lambda}(0) \psi_{\lambda}^\dagger(0) \bigg( 2 \pi \delta(\lambda) \bigg) \bigg] \rangle
\nonumber \\
& =  &
- \pi \langle  {\rm Tr} [ J_{\lambda=0}(0,0) ] \rangle .
\label{BC-identity2}
\end{eqnarray}
The identity that we will be using extensively is
\begin{eqnarray}
( x| \frac{1}{D \hspace*{-.2cm}\slash +m }|y)
\stackrel{m\rightarrow 0}{\longrightarrow}
\bigg[ \sum_{\lambda} \psi_{\lambda}(x) \psi_{\lambda}^\dagger(y)| \bigg( i {\rm P}(\frac{1}{\lambda})+  \pi \delta(\lambda) \bigg) \bigg],
\label{BC-identity3}
\end{eqnarray}
and
\begin{eqnarray}
\bigg[ ( x| \frac{1}{D \hspace*{-.2cm}\slash +m }|y) -i \gamma^5 ( x| \frac{1}{D \hspace*{-.2cm}\slash +m }|y)  i \gamma^5 \bigg]   &
\stackrel{m\rightarrow 0}{\longrightarrow} &
\bigg[ \sum_{\lambda} \psi_{\lambda}(x) \psi_{\lambda}^\dagger(y) \bigg( 2 \pi \delta(\lambda) \bigg) \bigg] \nonumber \\
& = &  \bigg[ J_{\lambda=0}(x,y) \bigg( 2 \pi  \bigg) \bigg]  .
\label{BC-identity4}
\end{eqnarray}
This relation itself is not gauge invariant, but still provides the essential intermediate step that relates chiral symmetry breaking to correlation functions.

\subsection{Chiral symmetry breaking}

As can be seen in Eq.~(\ref{BC-identity4}), the essence of chiral symmetry breaking is the non vanishing current density matrix of zero eigenvalues $J_{\lambda=0}(x,y)$.    That is when chiral symmetry is restored, any expectation value proportional to the density matrix will vanish.  For example, the disconnected diagram of a two point function is proportional to the following trace:
\begin{eqnarray}
{\rm Tr} [ ( x| \frac{1}{D \hspace*{-.2cm}\slash +m }|x) \Gamma] ={\rm Tr} [J_{\lambda=0}(x,x) \Gamma ],  \label{disconnected-1}
\end{eqnarray}
where $\Gamma$ is any gamma matrix coming from the current of the two point function $j^\Gamma=\bar{q} \Gamma q$.

\subsubsection{Gluon condensate and eigenvalues}

Some insights can be obtained by looking at the heavy quark condensate, which in the heavy quark mass expansion can be related to the gluon condensate.
\begin{eqnarray}
\langle \bar{h} h \rangle & = & -\frac{1}{m_h} \langle \frac{\alpha_s}{12 \pi} G^2 \rangle
\label{h-cond}
\end{eqnarray}

\begin{eqnarray}
\langle \bar{h} h \rangle & = &
-\frac{1}{V} \int d^4 x \langle {\rm Tr} [ ( x| \frac{1}{D \hspace*{-.2cm}\slash +m_h }|x) ] \rangle  \nonumber \\
&=& -\int d^4x \langle \sum_\lambda  \frac{\psi_\lambda(x)^\dagger \psi_\lambda(x) }{V} \frac{1}{m-i\lambda} \rangle \nonumber\\
& \stackrel{m_h\rightarrow \infty }{\longrightarrow} &
-\frac{1}{m} \int d^4x \langle \sum_\lambda  \frac{\psi_\lambda(x)^\dagger \psi_\lambda(x) }{V}  \rangle .
\label{h-cond2}
\end{eqnarray}
Therefore,
\begin{eqnarray}
\langle \frac{\alpha_s}{\pi} G^2 \rangle
=12 \langle \sum_\lambda  \rho(\lambda) \rangle .
\label{gl-cond}
\end{eqnarray}

\section{Topological configurations}

The integration over gauge fields given in Eq.(\ref{measure0}) can have nontrivial gauge configurations characterized by $\nu=(g^2/32 \pi^2) \int d^4x F \tilde{F} \neq 0$.  These configurations have $n_+$ ($n_-$) number of right handed (left handed) zero mode solutions that satisfy the index theorem $\nu=n_+-n_-$.  For these configurations it is useful to explicitly write the zero mode contribution as a separate quark determinant.  For example, for configurations with $\nu=1$,
\begin{eqnarray}
Z_{\nu=1} & = & \int [dA_\mu(x)]_{\nu=1}  \exp [-S_{YM} ]
[d q(x) d \bar{q}(x)] [dq_R d \bar{q}_L]  \exp [ -\int d^4x \bar{q}(x)[D \hspace*{-.2cm}\slash +m]q(x) ] \nonumber \\
& = &
\int [dA_\mu(x)]_{\nu=1}  \exp [- \int \frac{1}{2g^2} F^2]
{\rm det}' [D \hspace*{-.2cm}\slash +m] \times {\rm det} \big( \int d^4x \bar{\psi}_0(x) m \psi_0(x) \bigg)  \nonumber \\
& = & \int [d \mu]_{\nu=1}{\rm det} \big( \int d^4x \bar{\psi}_0(x) m \psi_0(x) \bigg) \label{measure11}
\end{eqnarray}
Here, $\psi_0$ is the zero mode solution and the measure in the zero mode runs over all flavors.  $\psi_0$ is the zero mode solution.  One should note that the contribution from these configurations is zero in the chiral limit.  However calculating the $n$-point functions involving quark operators with asymmetric chirality containing all flavors, the quark determinant from the zero modes does not appear and topological configuration will give non vanishing contributions.

\section{Adding the $\theta$ term}

Consider adding the $\theta$ term to the QCD action.
\begin{eqnarray}
Z_\theta & = & \int [dA_\mu(x)][dq(x) d\bar{q}(x)] \nonumber \\
 && \times \exp \bigg[-\int d^4x_E \bigg( \frac{1}{4g^2} F^2
+\bar{q}(  \gamma_\mu^ED_\mu+m_q) q + \theta \frac{i }{32 \pi^2} F \tilde{F} \bigg) \bigg]. \label{z-theta}
\end{eqnarray}

For our purpose, let us assume only one heavy quark with mass $m_Q$.  Then, under a chiral transformation $\theta \rightarrow \theta+ \pi/2$, the action will be invariant except for the following replacement:
\begin{eqnarray}
m_Q \rightarrow m_Q e^{i \gamma_5 \pi/2}=m_Q i \gamma_5.
\end{eqnarray}

\section{Heavy quark expansion}

Here, we rederive the heavy quark expansion for the lowest quark condensate\cite{Generalis:1983hb}.
In the fixed point gauge, the heavy quark propagator can be written as an expansion in gluon operators.  The term in the momentum space propagator that gives a gauge invariant term in the gauge invariant quark condensate operators is the following\cite{Kim:2000kj}.
\begin{eqnarray}
iS_{G^2}(p) & = & - \frac{i}{4}G_{\alpha \beta} G_{\rho \sigma} \bigg[
[\alpha, \beta, \rho, \sigma]+ [\alpha, \rho, \beta,  \sigma]+ [\rho, \alpha, \beta, \sigma] \bigg] \label{g2}
\end{eqnarray}
where,
\begin{eqnarray}
[\alpha, \beta, \rho, \sigma] = \frac{1}{ p \hspace*{-.13cm}\slash +m} \gamma_\alpha \frac{1}{ p \hspace*{-.13cm}\slash +m} \gamma_\beta \frac{1}{ p \hspace*{-.13cm}\slash +m} \gamma_\rho \frac{1}{ p \hspace*{-.13cm}\slash +m} \gamma_\sigma \frac{1}{ p \hspace*{-.13cm}\slash +m} .
\end{eqnarray}
Substituting,
\begin{eqnarray}
\langle \bar{q} q \rangle & = & - \lim_{x \rightarrow 0} {\rm Tr} \langle  {\cal T} \bigg( q(x) \bar{q}(0) \bigg) \rangle \nonumber \\
& = & - \int \frac{d^4 p}{(2 \pi)^4}  \langle {\rm Tr} [ iS(p)  ] \rangle.
\end{eqnarray}
Substituting Eq.~(\ref{g2}) into the above formula, one finds the following leading term in the heavy quark mass expansion for the heavy quark condensate.
\begin{eqnarray}
\langle m_h \bar{h}  h  \rangle  & = & -\frac{1}{12} \langle \frac{\alpha_s}{\pi} G^2 \rangle + O(\frac{1}{m_h^2})  \nonumber \\
\langle m_h \bar{h} i \gamma_5 h  \rangle  & = & -\frac{1}{8} \langle \frac{\alpha_s}{\pi} G \tilde{G}  \rangle + O(\frac{1}{m_h^2})  \label{heavy-q}
\end{eqnarray}

\end{document}